\documentclass[12pt]{article}
\usepackage[dvips]{graphics}
\usepackage{graphicx}
\title{Making nonlocal reality compatible with relativity}
\author{Hrvoje Nikoli\'c \\
Theoretical Physics Division, Rudjer Bo\v{s}kovi\'{c} Institute, \\
P.O.B. 180, HR-10002 Zagreb, Croatia \\
{\normalsize e-mail: hrvoje@thphys.irb.hr} \\
\makebox[1in]{} \\
}
\date{\today}
\begin{document}
\maketitle


\begin{abstract}
It is often argued that hypothetic nonlocal reality responsible
for nonlocal quantum correlations between entangled particles
cannot be consistent with relativity. I review the most frequent
arguments of that sort, explain how they can all be circumvented, and present 
an explicit Bohmian model of nonlocal reality (compatible with quantum phenomena)
that fully obeys the principle of relativistic covariance and does not
involve a preferred Lorentz frame.
\end{abstract}
\vspace*{0.5cm}
\maketitle

\section{Introduction}

The Bell theorem \cite{bell} shows that quantum mechanics (QM)
is not compatible with local reality. This suggests that reality might be nonlocal.
The best known model of nonlocal reality compatible with QM is provided by the Bohmian  
interpretation \cite{bohm1,bohm2} of nonrelativistic QM.
However, this particular model of nonlocal reality is not relativistic covariant.
In this sense, this model is not compatible with relativity.
Moreover, it is often argued that no model of nonlocal reality can be compatible
with relativity. Or more succinctly, that nonlocality and relativity cannot live together.
Thus, many believe that the whole idea of making nonlocal reality compatible with relativity
is an impossible task. 

Recently, however, an explicit relativistic-covariant version of the
Bohmian interpretation of QM has been introduced in \cite{nikijqi},
with some further developments presented in \cite{nikfp} and \cite{nikijmpa}.
This provides an explicit counterexample to various arguments that such a theory
should be impossible. But how that can be? Where do exactly the standard impossibility 
arguments fail? 
The purpose and intention of the present paper is to answer these questions and to
make this relativistic-covariant 
model of nonlocal reality understandable to a general physics audience.
With this intention in mind, in this paper I review some of the main ideas
presented in \cite{nikijqi,nikfp,nikijmpa}, and present some novel
qualitative and quantitative insights that make the whole idea 
conceptually clearer.
Sec.~\ref{CONC} is devoted to a qualitative non-technical explanation of the main ideas,
while the essential technical details of the relativistic-covariant theory are presented 
in Sec.~\ref{TECH}. 

\section{Conceptual issues: Frequent objections and responses}
\label{CONC}

Basically, the idea of nonlocal reality asserts
that nonlocal correlations between entangled particles
are caused by superluminal (i.e., faster than light) influences between the 
entangled particles. However, 
it is often argued that superluminal influences are in contradiction with 
the theory of relativity. In this section I review the most frequent arguments
for this contradiction and qualitatively explain how all these arguments can be easily circumvented.
I present it as a series of objections (O) and responses (R) having a form of a
dialogue.

O: The theory of relativity implies that nothing can travel faster than light.

R: No, the theory of relativity does not imply that. The best known counterexample
is a {\em tachyon}, hypothetical particle with mass squared $m^2<0$. 
It is a completely relativistic object, and yet it travels only faster than light.

O: A tachyon is a purely mathematical construct. There is no evidence that it exists
in Nature.

R: Maybe it exists but we have just not yet discovered it. But that's not what
I am trying to convince you. At the moment, my only point is that such an object, if exists, 
is compatible with relativity. 

O: OK, traveling faster than light may be compatible with relativity, 
but it leads to logical paradoxes. If a signal travels faster than light, then there is
a Lorentz frame in which it travels backwards in time. Then I can send a message
to the past, which may change the past making it incompatible with the
presence. For instance, I can send a signal that would cause killing of my grandfather,
which would be incompatible with my own existence. 

R: You cannot do that if these superluminal signals obey deterministic laws.
Namely, if nature is deterministic, then you don't have free will to choose
to send a signal as you wish. Instead, you can only send the predetermined signals
which are consistent with all already known facts about the presence and the past.

O: But I {\em do} have free will.

R: Actually, you have the {\em impression} of having free will. But this may be an illusion.
You cannot be aware of all processes in your brain and body. 
The events determined by causes which you are not aware of may be interpreted 
by your consciousness as being determined by free will, even if the true free will does not exist.

O: OK, maybe I don't have a true free will, but at least I have a {\em very persuasive}
illusion of free will. For all practical purposes this effective free will 
cannot be distinguished from a true one.
And from all my experience I know that this effective free will may influence the future
but not the past. Isn't that an evidence that signals to the past do not exist?

R: Well, it may be that superluminal signals are of a different nature than ordinary signals,
such that humans simply cannot be aware of the existence of superluminal signals.

O: Isn't it quite an unnatural assumption?

R: Actually, it is quite natural. Humans are macroscopic beings who perceive the world
in terms of classical phenomena. There is a lot of evidence, 
especially from the theory of decoherence \cite{decohbook,schloss}, that macroscopic classical physics 
emerges from the fundamental microscopic quantum physics.
By assumption, superluminal signals are inherently quantum phenomena responsible for nonlocal
correlations between entangled particles, so it is quite natural to expect that
their effect cannot be seen at the classical macroscopic
level at which decoherence effectively destroys the quantum correlations.

O: I am still not convinced. But to make the discussion clearer, let us avoid any mentioning
of humans, free will, and other vague stuff that are not well understood in physical terms.

R: I agree, let us avoid it.

O: So let as assume that there is a machine programmed such that it sends a message to the
past that commands its own destruction when this message is received. To be more specific, let ...

R: Stop, don't even bother with the details! It cannot work because a machine is a macroscopic
classical object, and I have already explained that superluminal signals do not work at the macroscopic
classical level.

O: OK, so let us replace the ``machine'' with a microscopic object (consisting of 
a few particles only) that behaves quantum mechanically.

R: A microscopic object cannot send a message that would contradict
its own existence.

O: Why not? 

R: First, because I assume that the microscopic object does not have free will, 
or even an illusion of free will, to send any message it ``wishes''. 
Second, even if I discard this assumption, I 
certainly {\em must} assume that the microscopic laws are self-consistent,
i.e., that such inconsistent systems do not appear as solutions of the mathematical
equations describing the microscopic laws.

O: Are you saying that you actually {\em can} construct such consistent microscopic laws?

R: Yes, you will see the details in the technical part of this paper.

O: OK, maybe you can achieve consistency with superluminal communication, but let me attack the whole issue from another side. 
If communication is superluminal, then there is a Lorentz frame in which it is
instantaneous. If the communication is instantaneous in one Lorentz frame, then it is not instantaneous in any other Lorentz frame. Therefore, there is a preferred Lorentz frame with respect to which the communication is instantaneous. Consequently, the principle of relativity is violated.

R: This is like using the following argument on {\em subluminal} communication.
If communication is subluminal, then there is a Lorentz frame in which the carrier of the message
is at rest. If it is at rest in one Lorentz frame, then it is not at rest in any other Lorentz frame. 
Therefore, there is a preferred Lorentz frame with respect to which the carrier is at rest. Consequently, the principle of relativity is violated.

O: But your argument is incorrect. It is the general law of motion that must have the same
form in any Lorentz frame. A particular solution (a particle at rest with respect to some particular
Lorentz frame) does not need to have the same form in all Lorentz frames.

R: Exactly! But you should realize that my incorrect argument is completely analogous
to yours. In other words, your argument is incorrect for exactly the same reason.
A particular solution (communication instantaneous with respect to some particular
Lorentz frame) does not need to have the same form in all Lorentz frames.

O: This analogy works if you exchange the roles of time and space. And you will
probably say that such an exchange should be allowed in a relativistic-covariant theory.

R: Exactly!

O: But is it compatible with the principle of causality?

R: It depends on what exactly do you mean by principle of causality.
It is compatible with determinism, i.e., with a possibility that all events are 
caused by some ``prior'' events. However, due to the superluminal influences,
``prior'' does not allways need to mean ``at an earlier time''.

O: But we know that nature is causal in this {\em latter} sense, in which ``prior'' does mean ``at an earlier time''.

R: We know that nature is causal in this sense at the classical macroscopic level,
at which, as I already explained, superluminal influences do not exist.
This type of causality may be violated at the microscopic level.

O: But we know it isn't. Relativistic quantum field theory (QFT) is a well-tested microscopic theory, causal in the sense that field operators commute at spacelike distances.
This is a manifestation of the well-known fact that relativistic QFT is a local theory.

R: QFT is indeed well-tested, but it contains both local and nonlocal properties.
In particular, QFT predicts nonlocal correlations between entangled particles,
and they have been observed in many experiments (see, e.g., \cite{aspect}). 

O: These nonlocal correlations cannot be used for superluminal signalling. 

R: That is true, but it does not imply that nonlocal correlations are not caused 
by superluminal influences. It is possible that superluminal influences exist at the microscopic level,
but that they cannot be controlled at the classical macroscopic level.
Of course, QFT alone with its standard purely probabilistic interpretation certainly does 
{\em not describe} such superluminal influences, but it does not exclude their existence
either (unless, of course, you assume that QFT with its standard interpretation
is the ultimate theory of everything).




O: OK, it seems that relativistic superluminal influences cannot easily be disproved by general
arguments.
Perhaps you can really construct an explicit theory satisfying all these features. 
(I need to see the technical part of the paper.) But even if you can, it seems to me that
a theory having all these features must be extremely unnatural and contrived. 
Can your theory be derived from some simple natural principles?

R: I'm glad that you asked it, because the most remarkable 
part of the theory is the fact that it follows from some rather simple and natural
principles. 

O: What these principles are? 

R: There are four of them: \\
1) Take the laws of physics seriously! \\ 
2) Take spacetime seriously! \\
3) Take relativistic wave equations seriously! \\ 
4) Take particles seriously!

O: I must admit, these principles look simple and natural. But they also look 
somewhat vague. How they lead to all these nontrivial features needed for 
compatibility between superluminal influences and relativity?

R: They are only the guiding conceptual principles, not mathematical principles, which is why
they are vague. But let me explain in more detail what I mean. \\
By 1) I mean that {\em everything}, including the human brain, obeys the physical laws. 
They will turn out to be deterministic laws, which excludes the existence of free will.
(In fact, probabilistic laws also exclude the existence of free will, but it is less obvious,
and we shall not need it.)  \\
By 2) I mean that time and space should be treated on an equal footing. Note
that in the usual formulation of QM, time and space are not treated on an equal footing.
First, for one particle described by the wave function $\psi({\bf x},t)$, the infinitesimal probability
in the usual formulation is $|\psi|^2d^3x$, 
while from a symmetric treatment of time and space one expects
$|\psi|^2 d^3x\,dt$.
Second, for $n$ particles the wave function in the usual formulation takes the form
$\psi({\bf x}_1,\ldots,{\bf x}_n,t)$, while from a symmetric treatment of time and space one expects
$\psi({\bf x}_1,t_1,\ldots,{\bf x}_n,t_n)$. 
I formulate QM such that fundamental axioms involve the expressions above in which
time and space are treated symmetrically,
and show that the usual formulation corresponds to a special case. \\
By 3) I mean that relativistic wave functions represent something real, and that
the wave functions {\em allways} obey their wave equations.
It implies that there is no collapse. \\
By 4) I mean that particles are pointlike objects that exist even when you don't measure them.
A combination of this with 1), 2) and 3) above naturally leads to a relativistic covariant version of the Bohmian interpretation of QM. It is both relativistic covariant and nonlocal essentially
because the particles are guided by wave functions which are both relativistic invariant and nonlocal.

O: Isn't it shown that the Bohmian interpretation requires a preferred Lorentz frame?

R: That is true in the usual formulation of the Bohmian interpretation based on the 
usual formulation of QM in which time and space are not treated on an equal footing.
When QM is generalized as outlined in 2) above, then the corresponding Bohmian
interpretation does not longer require a preferred Lorentz frame.

O: I think I've got a general idea now. But I'll not be convinced until
I see the technical details.

\section{Technical details}
\label{TECH}

\subsection{Relativistic probabilistic interpretation}

We use the relativistic notation $x=\{x^{\mu}\}=(x^0,x^1,x^2,x^3)$, 
where $x^0\equiv t$ is the time coordinate and $x^i$, $i=1,2,3$, are the space coordinates.
(The set of 3 space coordinates is also denoted by ${\bf x}$.)
As usual in relativistic QM, we work in units $\hbar=c=1$.
Thus, a 1-particle wave function can be denoted by $\psi(x)$. It is natural to introduce
the spacetime scalar product
\begin{equation}
 \langle\psi|\psi'\rangle = \int d^4x \, \psi^*(x)\psi'(x) ,
\end{equation}
and to normalize $\psi$ so that 
\begin{equation}
\langle\psi|\psi\rangle =1 .
\end{equation}
Then the quantity
\begin{equation}\label{dP}
dP=|\psi(x)|^2d^4x 
\end{equation}
is naturally interpreted as probability that the particle will be found in the 
(infinitesimal) 4-volume $d^4x$. 

At first sight, (\ref{dP}) is not compatible with the usual probabilistic
interpretation in the 3-space
\begin{equation}\label{dP3}
dP_{(3)} \propto |\psi({\bf x},t)|^2d^3x .
\end{equation}
Nevertheless, (\ref{dP}) and (\ref{dP3}) are compatible.
If (\ref{dP}) is the fundamental {\it a priori} probability, then (\ref{dP3})
can be interpreted as the {\em conditional} probability,
for the case in which one knows that the particle is detected at time $t$.
More precisely, 
\begin{equation}\label{5}
 dP_{(3)}=\frac{|\psi({\bf x},t)|^2 d^3x}{N_t} ,
\end{equation}
where
\begin{equation}\label{6}
N_t=\int d^3x |\psi({\bf x},t)|^2
\end{equation}
is the normalization factor.
Since $\psi$ is normalized so that $\int d^4x |\psi|^2=1$, we see that $N_t$
is also the marginal probability that the particle will be found at time $t$.

Can the probabilistic interpretation (\ref{dP}) be verified experimentally?
In fact, it already is.
Namely, in practice one often measures scattering cross sections or
decay widths and lifetimes associated with spontaneous decays 
of unstable systems.
The experimental results are in agreement with the standard theoretical predictions. 
My point is that these standard theoretical predictions actually {\em use}
(\ref{dP}), although not explicitly.
Let me explain it. 
In practice one calculates the transition amplitude $A$, which is 
the wave function at $t\rightarrow\infty$,
calculated under the assumption that the wave function at $t\rightarrow -\infty$ is known.
The energy conservation implies
\begin{equation}
A\propto\delta(E_{\rm in}-E_{\rm fin}) .
\end{equation}
Then the transition probability is proportional to
\begin{equation}\label{8}
|A|^2\propto[\delta(E_{\rm in}-E_{\rm fin})]^2= \frac{T}{2\pi}\delta(E_{\rm in}-E_{\rm fin}) ,
\end{equation}
where
\begin{equation}
T=\int dt =2\pi \delta(E=0) .
\end{equation}
However, $T$ is infinite, so the transition probability ({\ref{8}) does not make sense.
The standard reinterpretation is that the physical quantity is 
$|A|^2/T$, which describes the transition probability {\em per unit time}.
But this is equivalent to (\ref{dP}), which says that 
$\int d^3x |\psi|^2$ is not probability itself, but probability {\em per unit time}.
Even though the interpretation of $|A|^2/T$ as probability per unit time
may seem plausible even without the axiom (\ref{dP}), such an
interpretation is better founded in axioms of QM if (\ref{dP}) is also
accepted as one of the axioms.

Now let us generalize it to the case on $n$ particles.
Each particle has its own space position 
${\bf x}_a$, $a=1,\ldots,n$, as well as its own time coordinate $t_a$. 
The wave function has the form
$\psi(x_1,\ldots,x_n)$.
This is the so-called many-time wave function \cite{tomon}.
Now the fundamental probability is given by a generalization of (\ref{dP})
\begin{equation}\label{dPn}
dP=|\psi(x_1,\ldots,x_n)|^2 d^4x_1 \cdots d^4x_n .
\end{equation}
In particular, if the first particle is detected at time $t_1$, second particle at time $t_2$, etc., 
then the corresponding conditional probability is given by a generalization of (\ref{5})-(\ref{6})
\begin{equation}\label{tor11}
 dP_{(3n)}=\frac{|\psi({\bf x}_1,t_1,\ldots,{\bf x}_n,t_n)|^2 d^3x_1 \cdots d^3x_n}
{N_{t_1,\ldots,t_n}} ,
\end{equation}
\begin{equation}\label{tor12}
 N_{t_1,\ldots,t_n}=\int |\psi({\bf x}_1,t_1,\ldots,{\bf x}_n,t_n)|^2 d^3x_1 \cdots d^3x_n .
\end{equation}
The usual single-time probabilistic interpretation is recovered by taking the special case
$ t_1=\cdots=t_n\equiv t $ in (\ref{tor11})-(\ref{tor12}) .

\subsection{Quantum theory of measurements}

Let $\psi(x)$ be expanded as
\begin{equation}\label{nikolic:meas0}
 \psi(x)=\sum_b c_b \psi_b(x) ,
\end{equation}
where $\psi_b(x)$ are eigenstates of some hermitian operator $\hat{B}$ on the Hilbert space
of functions of $x$. Let $\psi_b(x)$ be normalized such that 
$\int d^4x \, \psi^*_b(x)\psi_b(x) =1$. Assume that one measures the value
of the observable $B$ described by the hermitian operator $\hat{B}$.
In a conventional approach to QM, one would postulate that $|c_b|^2$ is the probability that 
$B$ will take the value $b$. Nevertheless, there is no need for such a postulate
because, whatever the operator $\hat{B}$ is, this probabilistic rule can be derived
from the probabilistic interpretation in the position space (\ref{dPn}). 

To understand this, one needs to understand how a typical measuring apparatus works,
i.e., how the wave function of the measured system described by the coordinate $x$
interacts with the wave function of the measuring apparatus described by the coordinate $y$.
(For simplicity, we assume that $y$ is a coordinate of a single particle, but essentially
the same analysis can be given by considering a more realistic case in which
$y$ is replaced by a macroscopically large number $N$ of particles
$y_1,\ldots,y_N$ describing the macroscopic measuring apparatus. Similarly,
the same analysis can also be generalized to the case in which $x$ is replaced by
$x_1,\ldots,x_n$.)
Let the wave function of the measuring apparatus for times before the interaction
be $E_0(y)$. Thus, for times $x^0$ and $y^0$ before the interaction, the total wave function is
$\psi(x)E_0(y)$. But what happens after the interaction?
If $\psi(x)=\psi_b(x)$ before the interaction, then the interaction must be such that
after the interaction the total wave function takes the form $\psi_b(x)E_b(y)$, 
where $E_b(y)$ is a macroscopic state 
of the measuring apparatus, normalized so that $\int d^4y \, E^*_b(y) E_b(y) =1$.
The state $E_b(y)$ is such that
one can say that ``the measuring apparatus shows that the result of measurement is $b$''
when the measuring apparatus is found in that state.
Schematically, the result of interaction described above can be written as
\begin{equation}\label{nikolic:meas1}
\psi_b(x) E_0(y) \rightarrow \psi_b(x)E_b(y) .
\end{equation}
Of course, most interactions do not have the form (\ref{nikolic:meas1}), but only those
that do can be regarded as measurements of the observable $\hat{B}$.
The transition (\ref{nikolic:meas1}) is guided by some linear differential equation,
which means that the superposition principle is valid.
Therefore, (\ref{nikolic:meas1}) implies that for a general superposition (\ref{nikolic:meas0})
we have
\begin{equation}\label{nikolic:meas2}
\sum_b c_b \psi_b(x) E_0(y) \rightarrow \sum_b c_b \psi_b(x)E_b(y) \equiv \psi(x,y).
\end{equation}

The states $E_b(y)$ must be macroscopically distinguishable. In practice, it means
that they do not overlap (or more realistically that their overlap is negligible), i.e., that
\begin{equation}\label{nikolic:meas3}
E_b(y) E_{b'}(y) \simeq 0 \;\; {\rm for} \;\; b \neq b' ,
\end{equation}
for all values of $y$. Instead of asking ``what is the probability that the measured particle
is in the state $\psi_b(x)$'', the operationally more meaningfull question is
``what is the probability that the measuring apparatus will be found
in the state $E_b(y)$''. The (marginal) probability density for finding the particle describing 
the measuring apparatus at the position $y$ is
\begin{equation}\label{nikolic:meas4}
 \rho(y)=\int d^4x \, \psi^*(x,y)\psi(x,y) .
\end{equation}
Using (\ref{nikolic:meas2}) and (\ref{nikolic:meas3}), this becomes
\begin{equation}\label{nikolic:meas5}
 \rho(y) \simeq \sum_b |c_b|^2 |E_b(y)|^2 .
\end{equation}
Now let ${\rm supp}\, E_b$ be the support of $E_b(y)$, i.e., the region of 
$y$-space on which $E_b(y)$ is not negligible. 
Then, from (\ref{nikolic:meas5}), the probability that $y$ will take a value from the
support of $E_b(y)$ is
\begin{equation}\label{nikolic:meas6}
 p_b=\int_{{\rm supp}\,E_b} d^4y \, \rho(y) \simeq  |c_b|^2 .
\end{equation}
In other words, the probability that the measuring apparatus will be found
in the state $E_b(y)$ is (approximately) equal to $|c_b|^2$. 

\subsection{Wave equation and the Bohmian interpretation}

Now 
we are finally ready to study an explicit quantum relativistic model of nonlocal reality. 
As the simplest nontrivial example,
we study a system of $n$ spinless relativistic non-interacting 
(but possibly entangled) particles.
Their wave function satisfies the $n$-particle Klein-Gordon equation 
\begin{equation}\label{KGn}
\sum_{a=1}^{n} [\partial_a^{\mu}\partial_{a\mu}+m^2_a] \psi(x_1,\ldots,x_n) =0 ,
\end{equation}
where we use Minkowski metric with the signature $(+---)$.
We introduce the relativistic Bohmian interpretation, according to which 
each particle has its own trajectory $X^{\mu}_a(s)$ in spacetime,
where $s$ is an auxiliary scalar parameter.
The wave function $\psi(x_1,\ldots,x_n)$ does not depend on $s$. 
By writing the complex wave function in the polar form 
$\psi=|\psi|e^{iS}$, one finds that 
the Klein-Gordon equation (\ref{KGn}) implies a relativistic conservation equation
\begin{equation}\label{relcons}
 \frac{\partial |\psi|^2}{\partial s} + \sum_{a=1}^{n} \partial_{a\mu}(|\psi|^2 v^{\mu}_a) =0 ,
\end{equation}
where 
\begin{equation}
v^{\mu}_a(x_1,\ldots,x_n) \equiv -\partial_a^{\mu}S(x_1,\ldots,x_n) .
\end{equation}
This implies that it is consistent to postulate that the Bohmian 
trajectories are determined by equations
\begin{equation}\label{bohm}
\frac{dX^{\mu}_a(s)}{ds}=v^{\mu}_a(X_1(s),\ldots,X_n(s)) .
\end{equation}
Namely, if a statistical ensemble of particles has the distribution (\ref{dPn})
for some initial $s$, then the conservation equation (\ref{relcons}) provides that
the ensemble will have the distribution (\ref{dPn}) for any $s$.

We see that the equation of motion (\ref{bohm}) is {\em nonlocal}, because
the velocity of one particle for some value of $s$ depends on the 
positions of all other particles for the same value of $s$.
On the other hand, we also see that the theory is {\em relativistic covariant},
because no {\it a priori} preferred  coordinate frame is involved.
Both the equations of motion (\ref{KGn}) and (\ref{bohm}) and the 
probabilistic interpretation (\ref{dPn}) have the same form in all
Lorentz frames.

As we have explained, (\ref{relcons}) implies that
particles have the same distribution of spacetime positions as
predicted by the purely probabilistic interpretation (\ref{dPn}). 
But what about other measurable quantities? For example,
what about the space distribution of particles described in purely probabilistic QM by (\ref{dP3})?
Or what about the statistical distribution of particle velocities? In general, 
in the Bohmian interpretation all these other quantities may have a distribution totally
different from those predicted by purely probabilistic QM. In particular, the Bohmian velocities
of particles may exceed the velocity of light, 
while purely probabilistic QM does not allow such velocities
because the eigenstates $e^{-ip_{\mu}x^{\mu}}$ of the velocity operator
$i\partial_{\mu}/m$ are not solutions of the Klein-Gordon equation for $p^{\mu}p_{\mu}<0$.
Yet, when a quantity is {\em measured}, then the two theories have {\em the same} 
measurable predictions. 
Namely, since the Bohmian interpretation is compatible with (\ref{dPn}), the probability
that the measuring apparatus will be found in the state $E_b(y)$ in (\ref{nikolic:meas2})
is given by (\ref{nikolic:meas6}), which is the same as that in the purely probabilistic interpretation.

\section*{Acknowledgements}

This work was supported by the Ministry of Science of the
Republic of Croatia under Contract No.~098-0982930-2864.

\end{document}